\documentclass[11pt]{IEEEtran}
\usepackage{csquotes}
\usepackage{balance}

\usepackage[hyphens]{url}
\usepackage{hyperref,csquotes}

\usepackage{graphicx}
\graphicspath{{images/}}
\DeclareGraphicsExtensions{.png}

\usepackage{float}
\usepackage[labelfont=bf]{caption}
\usepackage[many]{tcolorbox}

\usepackage{array}

\usepackage[super]{natbib}

\newtcolorbox[blend into=figures]{myfigure}[2][]{float*=htb,capture=hbox,
	blend before title code={\fbox{##1}\ },	title={#2},
	every float=\centering,#1}

\definecolor{captioncol}{RGB}{210,35,43}
\definecolor{backcol}{RGB}{246,241,235}

\newtcolorbox{mycaptionbox}{
	freelance,
	enhanced,
	colframe=gray,
	arc=0pt,
	middle=1pt,
	outer arc=0pt,
	boxrule=1pt,
	enlarge left by=-5mm,
	enlarge right by=-5mm,
	text width=\textwidth-1.4pt,
	nobeforeafter,
	segmentation hidden=true,
	bottom=-1.5ex,
	interior code={
		\path[draw=none,fill=backcol!30]
		(interior.south west) rectangle (segmentation.north east);
		\path[draw=none,fill=white]
		(segmentation.west) rectangle (interior.north east);
		\draw[draw=gray,line width=1pt]
		(segmentation.west) -- (segmentation.east);
	},
}

\newenvironment{figureboxstar}[1][tbp]
{\begin{figure*}[#1]\begin{mycaptionbox}}
		{\end{mycaptionbox}\end{figure*}}

\DeclareCaptionFormat{colorlabel}{\textcolor{captioncol}{#1#2}#3}
\newcommand{\mysection}[1] 
{
	\noindent\rule{\linewidth}{0.8pt}\\[2pt]
	\centerline{#1}\\[-6pt]
	\noindent\rule{\linewidth}{0.8pt}\\[-8pt]
}

\makeatletter

\renewcommand{\section}{%
	\@ifstar{%
		\mysection%
	}{%
		\mysection%
	}%
}

\makeatother

\begin{document}
\onecolumn

\title{COVID-19 and the Social Distancing Paradox: dangers and solutions}
\author{\IEEEauthorblockN{Massimo Marchiori$^{1,2}$}\\
	\IEEEauthorblockA{$^1$University of Padua\\
		Via Trieste 63, 35121 Padova, Italy \\}
	\IEEEauthorblockA{$^2$European Institute for Science, Media and Democracy\\
		Boulevard Louis Schmidt 24, 1040 Brussels, Belgium\\
		Email: massimo.marchiori@unipd.it}}
\maketitle

\section{ABSTRACT}
\\[5px]
\noindent
\normalsize 
\textbf{Background}\\
Without proven effect treatments and vaccines, Social Distancing is the key protection factor against COVID-19. Social distancing alone should have been enough to protect again the virus, yet things have gone very differently, with a big mismatch between theory and practice. What are the reasons? A big problem is that there is no actual social distancing data, and the corresponding people behavior in a pandemic is unknown. We collect the world-first dataset on social distancing during the COVID-19 outbreak, so to see for the first time how people really implement social distancing, identify dangers of the current situation, and find solutions against this and future pandemics.

\noindent
\textbf{Methods}\\
Using a sensor-based \enquote{social distancing belt} we collected social distance data from people in Italy for over two months during the most critical COVID-19 outbreak. Additionally, we investigated if and how wearing various Personal Protection Equipment, like masks, influences social distancing.

\noindent
\textbf{Results}\\
Without masks, people adopt a counter-intuitively dangerous strategy, a paradox that could explain the relative lack of effectiveness of social distancing.
Using masks radically changes the situation, breaking the paradoxical behavior and leading to a safe social distance behavior. In shortage of masks, DIY (Do It Yourself) masks can also be used: even without filtering protection, they provide social distancing protection. Goggles should be recommended for general use, as they give an extra powerful safety boost. Generic Public Health policies and media campaigns do not work well on social distancing: explicit focus on the behavioral problems of necessary mobility are needed.



\twocolumn
\section{INTRODUCTION}

The recent COVID-19 pandemic has caught many governments off-guard, not only for the rapid initial spread of the virus, but more importantly for the unexpected growth of the contagion, even in presence of safety protocols. One might wonder what are in fact the main reasons for such situation. On the one hand, we have various recommendations issued by the World Health Organization (WHO) and by local authorities to limit the spreading of the virus: such measures have grown tighter with the diffusion of the contagion. On the other hand, we have the companion power of prediction models that should have also given precise indications in this battle against the virus. Both of these actions, theory and practice combined, should have provided for a rapid containment of the virus and a quick return to a normal life.

In reality, things have gone in a very different way. The strict recommendations progressively enforced by authorities to limit sociality, most notably lock-down and social distancing, have in fact proved surprisingly ineffective when compared to their theoretical impact. A sufficiently large social distancing should alone be a formidable protection measure\cite{kelso_simulation_2009,ahmed_effectiveness_2018,ngonghala_mathematical_2020}, but this has not been the case, as shown by the contagion growth data in many countries.

Alongside, the mathematical models that should have offered guidance against the virus have faced big difficulties to provide precise predictions and consequent fine insights on how to shape the best containment strategy\cite{adam_special_2020,jewell_caution_2020}.

Why is it so? Regarding public health measures, obviously there has been some mismatch between theory and practice, between the dictated recommendations and the actual behavior of the people. This mismatch can also explain the difficulties encountered in designing models, given that precise data are essential to understand the logic governing the spreading of the virus, whereas so far we have been dealing with very limited datasets collected \emph{a posteriori}. For instance, COVID-19 data collected and used in Italy both by government and local authorities to make decisions, and by researchers to build predictive models, are accessible in a GitHub repository\cite{noauthor_covid-19_2020} containing classic a posteriori data like for example the number of people infected, hospitalized or dead in regions and provinces. These kinds of data are certainly essential and helpful, but they do not provide direct insights on the actual causes of the development. What we miss is precise data on the key behavior of people during the infection, so to understand the underlying mechanics, build causal models with very low error margins, and guide effective public health policies.

In this research we tackle this problem by addressing the key component of social distancing. For the first time, we collect real data on social distancing in a pandemic situation, analyze the actual shape of social distancing as performed by people, identify a paradoxical default behavior of social distancing that can explain the dangerous spread of COVID-19, and provide corresponding functional actions that can be taken to help against this and future pandemics.

\section{METHODS}

Starting from extensive experience in sensor-based equipment and smart city solutions (for instance in city environments\cite{Marchiori-FixMeUp,Marchiori-MindYourStep,Marchiori-SafeCycle,Marchiori-Health}, shopping malls\cite{Marchiori-MallParadox,Marchiori-BackToTheBazaar}, garbage collection\cite{Marchiori-SmartCheapCity,Marchiori-BinsWithEyes} and more\cite{Marchiori-Cows}), we have designed and built a sensor-equipped \emph{Social Distancing belt}, a special belt augmented with hidden sensor boxes allowing to measure social distances. The sensor boxes are concealed so to appear as normal belt pouches, therefore allowing for discrete and unnoticed operation. A social distancing box relies on a hardware core composed by an ATMEGA 328P compatible micro-controller, custom wired and endowed with Real Time Clock (RTC) capabilities via a DS1338 chip, MicroSD card module for storage, and an ultrasonic distance sensor HY-SRF05 operating at 40KhZ. Main power is provided by 20000mAh rechargeable power banks, whereas a common 3V CR2032 coin cell battery powers the RTC component. In order to avoid spurious readings, and rule out cases like for instance couples, parents with children, people with dogs and so on (all cases that obviously alter the social distancing behavior and available room) the system is in standby and social distancing measures can be activated and deactivated via an on/off switch. Temporal reasoning programmed in the micro-controller deals with cases of people quickly passing one by another (local minima in the social distance curve are extracted, using threshold values), so freeing the operator from dealing with continuous on/off cycles: during normal operation the switch is used only to rule out the unwanted cases (or optionally to save energy when no person is nearby). As said before, the whole assembly is concealed so no wiring or other electronic material actually shows up, so externally the apparatus looks like a normal belt with some pouches.

\begin{figureboxstar}[hbtp]
	\centering
	\includegraphics[width=0.45\textwidth]{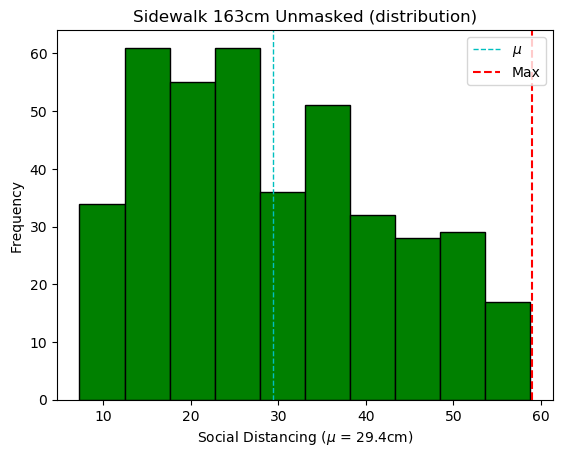}
	\tcblower
	\caption{\textbf{Paradoxical Social Distancing behavior without a mask during the COVID-19 pandemic.}\\Distribution histogram of social distancing in 163cm sidewalks. $\mu$ is the average social distance, Max is the maximum distance obtainable when staying within the sidewalk.}
	\label{fig:PATH-163-unmasked}
\end{figureboxstar}

Using the belt, we have proceeded to collect social distance information in the Venice metropolitan area (Italy) during the most critical period of the COVID-19 pandemic: data collection has taken place uninterruptedly for a period of over 2 months, from February 24 to April 29, 2020.

Even during the lock-down periods, people are still allowed to go out for essential purposes and consequent \emph{necessary mobility}, like going shopping for food. In Italy for instance using public transportation has been actively discouraged, asking people to walk instead. So, we have targeted focus points where social distancing is crucially important, and where data can be continuously collected even during the worst periods of a pandemic, like pedestrian sidewalks and food shops.

In this first preprint we present the result collected by monitoring sidewalks. In order to see what happens in various situations we have selected sidewalks of different widths: 163cm, 175cm and 222cm. The belt operator always stays on one side of the sidewalk, maximizing the distance from another person. All the selected sidewalks also allow to temporarily step out (being parallel to a green area or to a bicycle path): this way we can measure if the pressure of social distancing in the pandemic makes people gain further distance when approaching someone. In fact, in 167cm and 175cm sidewalks the only way to get the recommended minimum distance of 1 meter is to step out (or alternatively to slide sideways\cite{fruin_pedestrian_1987}).

For every chosen sidewalk we study social distancing in five different cases:
\begin{enumerate}
	\item \emph{Unmasked} case: the operator does not wear a mask
	\item \emph{Masked} case: the operator wears a surgical mask
	\item \emph{DIY-masked} case: the operator wears a DIY (Do It Yourself) mask
	\item \emph{Goggles masked} case: like the second (masked) case, with the operator additionally wearing goggles
	\item \emph{Goggles DIY-masked} case: like the third (DIY-masked) case, again with the operator additionally wearing goggles.
\end{enumerate}

The first case allows to determine the default behavior of people in a pandemic. The second and third case allows to check whether visual factors (wearing a mask) affect social distancing. Given the shortage of protective masks experienced in many countries, we also test DIY (Do It Yourself) masks, stressing their visual component: the mask has practically no protection value but it is purposely rough (made up by a piece of baking paper), so to be visually very noticeable as a bad home-made patch. The last two cases allow to verify whether goggles for eye protection (equipment not actually included in recommended public guidelines) further changes common social distancing behavior during a pandemic.

For the sake of readability, in this preprint we include graphics for the 163cm case, granted that the 175cm and 222cm cases exhibit the same behaviour.

\section{RESULTS}
\\[-2.6em]
\subsection*{The Unmasked Case}

Figure~\ref{fig:PATH-163-unmasked} shows the histogram distribution of social distances, helping to identify what kind of pattern people follow. The plot shows also the average social distance $\mu$ (in this case, 29.4cm) and the available Max space (maximum social distance available while staying on the sidewalk). We can see that in this case all people spatially distribute within the maximum width (so, not stepping out of the sidewalk).

The first important fact to notice is that the average social distance behavior does not work as we would expect during such a threatening pandemic. A safe social distance strategy would dictate people to use all the space at their disposal, therefore maximizing distance and having a distribution strongly skewed towards the Max. Instead, what happens is a completely paradoxical situation: not only people do not use this strategy, but actually follow an anomalous normal-like distribution which is skewed more towards the other person in the sidewalk (or in fact, towards the central area of the sidewalk). So, surprisingly, people tend to stay closer to another passing person rather than stay far away. This highly dangerous behavior is also consistent among different widths (175cm and 222cm).

This paradoxical behavior implies that requiring social distancing alone not only does not work well for sidewalks but actually maintains very dangerous proximity situations. Favoring walks instead of public transportation therefore does not  automatically produce a safer social environment.

\subsection*{The Masked Case}

The results of the second scenario (operator wearing a mask) are shown in the left panel of Figure~\ref{fig:MULTI-PATH-163-masks} for 163cm sidewalks. The introduction of a mask provokes a radical change of social distancing, making it grow (the average passes from 29.4cm to 58.42cm) and also modifying the overall distribution. Whereas in the unmasked case we had a distribution skewed towards the operator, in this scenario the paradox disappears, and the peak is instead farther away. Last but not least, we can observe an interesting phenomenon: the distribution actually extends beyond the Max limit, passing from 58.8cm of the unmasked case to 119.5cm. This behavior is consistent in all the other sidewalks (175cm and 222cm).

\begin{figureboxstar}[htbp]
	\centering
	\includegraphics[width=0.45\textwidth]{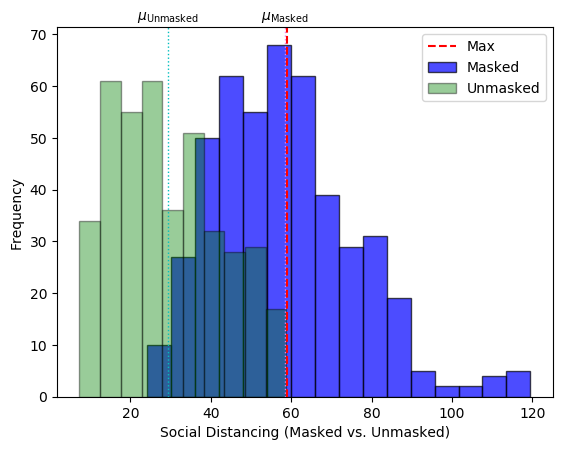}
	\hspace{1em}%
	\includegraphics[width=0.45\textwidth]{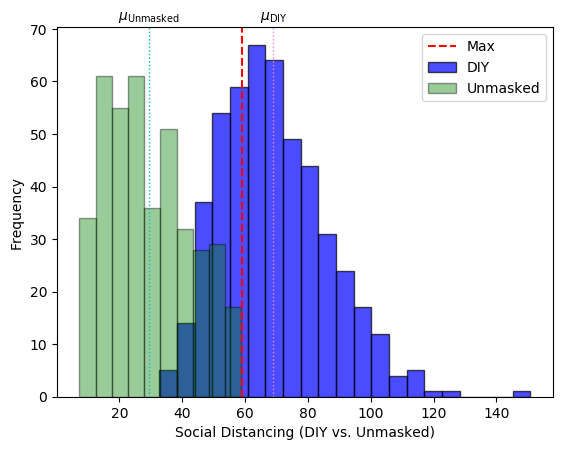}
	\tcblower
	\caption{\textbf{Social Distancing difference when wearing masks during the COVID-19 pandemic.}\\Left panel compares wearing a mask to the unmasked case, right panel wearing a DIY mask (163cm sidewalks).}
	\label{fig:MULTI-PATH-163-masks}
\end{figureboxstar}

\subsection*{The DIY-Masked Case}

Figure~\ref{fig:MULTI-PATH-163-masks} (right panel) shows what happens when wearing a DIY mask (163cm sidewalks). The situation is in all similar to the masked case: the distribution is again skewed in the same asymmetrical way, with people distancing beyond the sidewalk max width. The only notable difference is the average distance and overall width of the distribution, which grows even more than in the Masked case: average grows to 69.02cm and the whole distribution extends up to over 150cm of social distance. Again, this behavior is consistent among differently sized sidewalks (175cm and 222cm).

\subsection*{The Goggles Masked Case}

Figure~\ref{fig:MULTI-PATH-163-goggles} (left panel) shows what happens in 163cm sidewalks when we add goggles to masks. We can see that the effect is similar to what happened with DIY masks: the distribution skews further to bigger social distances. In fact, this extra distance boost is bigger than what obtained with a DIY mask, as the average social distance grows to 79.79cm. Again, the effect is consistent along all sidewalks (175cm and 222cm).

\begin{figureboxstar}[htbp]
	\centering
	\includegraphics[width=0.45\textwidth]{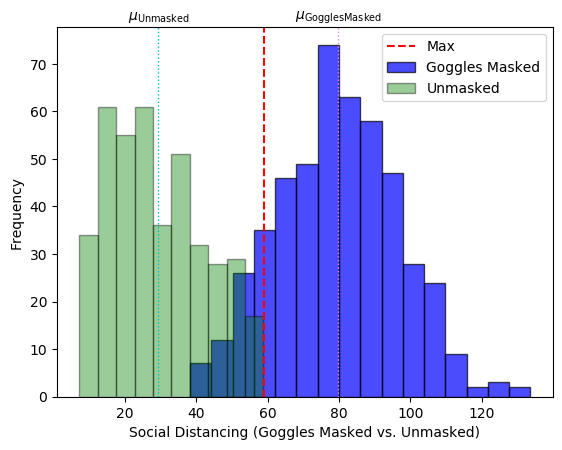}
	\hspace{1em}%
	\includegraphics[width=0.45\textwidth]{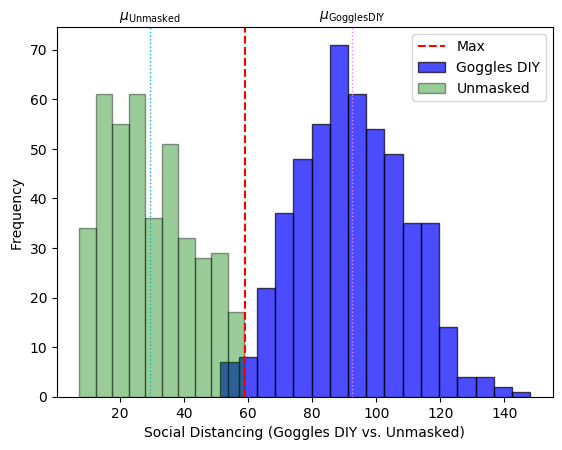}
	\tcblower
	\caption{\textbf{Social Distancing difference when wearing goggles and masks during the COVID-19 pandemic.}\\Left panel compares wearing goggles and mask to the unmasked case, right panel wearing goggles and a DIY mask (163cm sidewalks).}
	\label{fig:MULTI-PATH-163-goggles}
\end{figureboxstar}

\subsection*{The Goggles DIY-Masked Case}

The last case is the combination of goggles and DIY masks, shown in Figure~\ref{fig:MULTI-PATH-163-goggles} (right panel) for 163cm sidewalks. The effect is similar to the goggles masked case: adding goggles provides an extra social distance boost. Interestingly, this boost  is cumulatively added on top of that coming from the DIY mask. In the 163cm case the average social distance grows to 92.39cm and similar boosts are obtained for all the other sidewalks (175cm and 222cm).

\subsection*{Temporal Evolution during the COVID-19 Pandemic}

Alongside the spread of the virus, the Italian government issued a series of law decrees progressively setting more and more restrictions.

On the 30th of January 2020 WHO declared COVID-19 an international health emergency. The situation in Italy apparently remained good until late February when, given the outbreak of the virus, the Italian government reacted with some law decrees stating safety rules for the population. In parallel with the subsequent growth of the infection, the government (in a paradigmatic strategy of incremental national interventions followed by many other states) reacted with further decrees, progressively restricting social movements, aggregation places and workplaces. 
Among the various national decrees issued to prevent COVID-19, we can mark three main dates (dubbed N1, N2, N3 for later reference):
\begin{itemize}
	\item N1 (March 1): First social distancing enforcement\cite{noauthor_italian_2020} (dividing Italy into \enquote{red}, \enquote{yellow} and \enquote{red} zones). This measure implemented for the Veneto region a partial lockdown (schools closed, 1-meter social distancing in restaurants and bars)
	\item N2 (March 8): Partial lockdown extended to all Italy, 1-meter social distancing recommended also outdoors\cite{noauthor_italian_2020-1}
	\item N3 (March 11): Full lockdown\cite{noauthor_italian_2020-2}, only essential  movements allowed.
\end{itemize}

Additionally, on March 9 the government also launched a continuative massive media campaign dubbed \#Iorestoacasa (\enquote{\#Istayathome}), using celebrities from various fields to promote the public health recommendations by asking people to stay at home \cite{noauthor_italian_2020-3}.

The Veneto region has been an even more interesting case given that, besides national laws, the regional government has introduced unique additional safety measures. The most notable regional ordinances are three, setting additional public health rules in the following dates (dubbed R1, R2, R3):

\begin{itemize}
	\item R1 (March 21): Closing all shops on Sundays and enforcing a strict rule about the max distance (200 meters from home) allowed for walking without a compelling reason\cite{noauthor_veneto_2020}
	\item R2 (April 4): Compulsory masks in all shops\cite{noauthor_veneto_2020-2}
	\item R3 (April 14): Compulsory masks outdoor, end of the 200 meters distance restriction\cite{noauthor_veneto_2020-1}
\end{itemize}

The outcome of all these laws has been relatively disappointing: government restrictions and media outreach did not substantially change the situation, leading to a pandemic that has kept the whole Italian country stalled for months.

We now analyze how all these various health policies impacted social distancing.
Figure~\ref{fig:AVGS-Sidewalk-163} shows the temporal evolution of social distancing during the pandemic, allowing to see the effects of national decrees (N1, N2, N3),  the media campaign (M) and regional ordinances (R1, R2, R3). As we can see, there has been no noticeable difference: social distancing behavior over time has shown remarkable consistency. For instance, checking on the effect of the media campaign we find that in the unmasked case the average social distancing changed only by -1.8cm from the week before to the next week. Similar small differences also occurred in the masked (+3.6cm), DIY-masked (+1.6cm), goggles masked (+0.9cm) and goggles DIY-masked (+0.7cm) cases. All these small differences are compatible with normal data fluctuations, as substantiated by the Kolmogorov-Smirnov test: social distancing continues to follow the same distributions even after the campaign (the p-values for the five cases are 0.59, 0.82, 0.70, 0.93, 0.85 respectively).

The same situation occurs for all the other (national and regional) public health enforcements in all the other sidewalk cases (175cm and 222cm).

\begin{figureboxstar}
	\includegraphics[width=\textwidth]{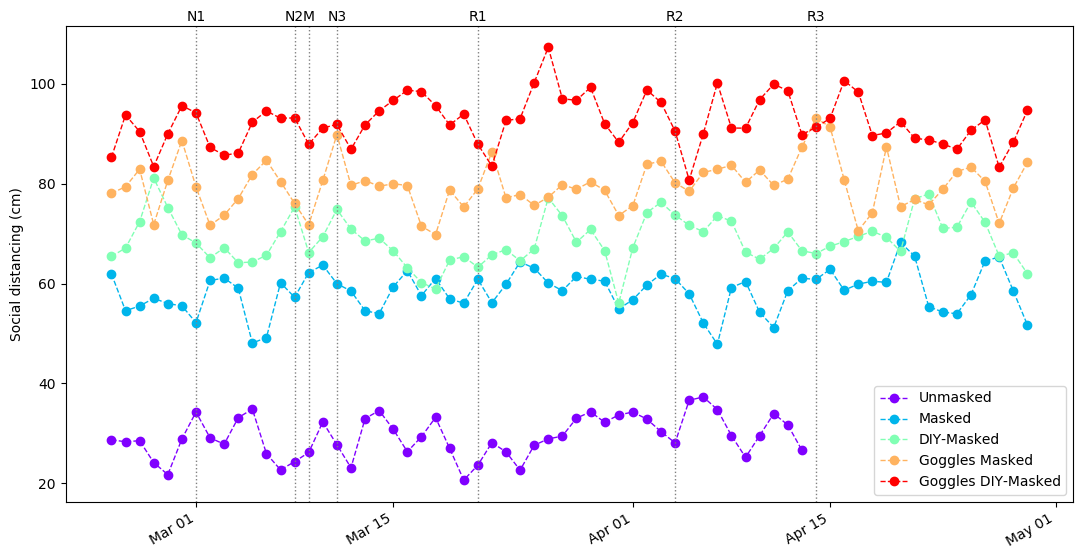}
	\tcblower
	\caption{\textbf{Temporal evolution of Social Distancing during the COVID-19 pandemic.}\\
		Comparison of the daily average social distancing among all the five scenarios from February 24 to April 29, 2020 (163cm sidewalks). N1, N2 N3 mark the national decrees, R1, R2, R3 the regional ordinances and M the start of the national media campaign. The unmasked case stops with R3 because of the obligation to wear masks outside.}
	\label{fig:AVGS-Sidewalk-163}
\end{figureboxstar}

\section{DISCUSSION}

The common behavior of Social Distancing exhibits a paradoxical and very dangerous behavior, which could explain the relative inefficiency of Social Distancing alone to disable the virus spreading. The essence of this paradox probably lies in the intrinsic social nature of our human collective, that social component that makes us aggregate and live in communities, so favoring social versus anti-social behavior. This built-in behavior is likely hard to alter, especially in everyday situations like walking: consequently, this counter-intuitive behavior of social distancing in the unmasked case is very dangerous and it should be subject of special attention, being a critical risk factor even in quarantine, due to necessary mobility (people have to go out at least to grab food). Limiting transportation mobility, like done in Italy, can in fact aggravate this problem. Even worst, this risky social behavior has been resilient to enforcement and media campaigns, thus also explaining the inefficiency of many public health measures against COVID-19.

A key strategy to get rid of the paradox and its danger is to try to de-activate the built-in social rules of people, by triggering an explicit counter-effect mechanism and so turning unconscious social behavior into conscious anti-social behavior. This safety trigger can be activated via visual stimuli that remind of the danger of social proximity, making people change their common dangerous behavior and implement a safer social distancing strategy.

Wearing a mask instead triggers this sort of repulsive effect, \enquote{pushing farther} people and the skew of the distribution, changing their common behavior so to gain social distance (even by stepping out of sidewalks). The distance increases with patched DIY masks, confirming the effect of a visual stimulus. Goggles act as \enquote{social distance boosters}, again consistently with the hypothesis that a visual stimulus signaling danger makes people more sensible to safety and increases social distancing protection. The good news is that the effectiveness of this booster seems also consistent with time.

Consequently, a number of actions can be suggested to help against the virus:

\begin{LaTeXdescription}
	\item[\normalfont\itshape Wearing Masks.] In the initial onset of the contagion, masks were not recommended in Italy but only to infected persons. Similar recommendations have been given by many other countries\cite{world_health_organization_advice_2020,leung_mass_2020, cdc_personal_2020,feng_rational_2020}.
	Given that masks lead people to implement social distancing in a safe way and also provide an overall distance boost, usage of masks should be in fact be always recommended to everyone in spite of their protection effectiveness\cite{greenhalgh_face_2020} and fitting troubles\cite{pauli_importance_2014}.
	\item[\normalfont\itshape Mask shortage and DIY.] One of the biggest problems many infected nations have been facing is availability of masks for all the population\cite{world_health_organization_shortage_nodate,sayburn_are_2020,us_food_and_drug_administration_faqs_2020}, given also their disposable nature. Lacking proper masks, DIY ones can profitably be used: even with no filtering protection, they provide social distancing protection. So, DIY masks can help a lot in all those situations where proper filtering masks are lacking or in shortage. And even when masks are available, adding a DIY mask on top still gives some extra protection.
	\item[\normalfont\itshape Goggles.] Goggles are currently not recommended protection equipment for the population, but they should, given they provide a significant additional boost to social distance.
	\item[\normalfont\itshape Social Distancing, Media and Education.] Suggesting or imposing Social Distancing, via media campaigns or laws, does not work so well on Social Distancing itself and leads to paradoxically dangerous situations. Therefore, Public Health campaigns should not focus only on limiting mobility, asking people to stay at home and providing generic social distancing rules, but also explicitly focus on the key aspects and behavioral problems of necessary mobility and of the dangerous paradox lying within our social nature.
\end{LaTeXdescription}

The analyzed social distancing behavior can be also used to obtain more precise prediction models (and so, better countermeasures): social distancing cannot be assumed as given, dictated by law or media, it's much more complex and also depends heavily on a so far neglected human factor, thus deserving specific attention  leading to a real \emph{social distancing science}.

Last but not least, the findings are not only local to the ongoing COVID-19 pandemics but are of general interest in future pandemic situations, given the general nature of the social distancing paradox, and the visual nature of social distancing boosters. 

\bibliographystyle{IEEEtran}

\balance

\bibliography{COVID,MM14}

\end{document}